\documentclass[aps,pra,twocolumn,showpacs,amsmath,amssymb,superscriptaddress,reprint]{revtex4-1}

\usepackage[utf8]{inputenc}

\usepackage{units}
\usepackage[colorlinks=true,citecolor=blue]{hyperref}
\usepackage{graphicx}

\usepackage{dcolumn}

\newcommand{\kBT}{k_\text{B}T}
\renewcommand{\vec}[1]{\mathbf{#1}}

\begin{document}
\title{Illumination-induced nonequilibrium charge states in self-assembled quantum dots}

\author{Sascha~R.~Valentin}
\affiliation{Lehrstuhl f\"ur Angewandte Festk\"orperphysik, Ruhr-Universit\"at Bochum, D-44780 Bochum, Germany}

\author{Jonathan~Schwinger}
\affiliation{Lehrstuhl f\"ur Angewandte Festk\"orperphysik, Ruhr-Universit\"at Bochum, D-44780 Bochum, Germany}

\author{Pia~Eickelmann}
\affiliation{Fakult\"at für Physik and CENIDE, Universit\"at Duisburg-Essen, Lotharstraße 1, 47048 Duisburg, Germany}
\affiliation{Lehrstuhl f\"ur Angewandte Festk\"orperphysik, Ruhr-Universit\"at Bochum, D-44780 Bochum, Germany}

\author{Patrick~A.~Labud}
\affiliation{Lehrstuhl f\"ur Angewandte Festk\"orperphysik, Ruhr-Universit\"at Bochum, D-44780 Bochum, Germany}

\author{Andreas~D.~Wieck}
\affiliation{Lehrstuhl f\"ur Angewandte Festk\"orperphysik, Ruhr-Universit\"at Bochum, D-44780 Bochum, Germany}

\author{Bj\"orn~Sothmann}
\affiliation{Fakult\"at für Physik and CENIDE, Universit\"at Duisburg-Essen, Lotharstraße 1, 47048 Duisburg, Germany}

\author{Arne~Ludwig}
\affiliation{Lehrstuhl f\"ur Angewandte Festk\"orperphysik, Ruhr-Universit\"at Bochum, D-44780 Bochum, Germany}

\date{\today}

\begin{abstract}
We report on capacitance-voltage spectroscopy of self-assembled InAs quantum dots under constant illumination.
Besides the electronic and excitonic charging peaks in the spectrum reported earlier, we find additional resonances associated with nonequilibrium state tunneling unseen in C(V) measurements before. 
We derive a master-equation based model to assign the corresponding quantum state tunneling to the observed peaks.
C(V) spectroscopy in a magnetic field is used to verify the model-assigned nonequilibrium peaks.
The model is able to quantitatively address various experimental findings in C(V) spectroscopy of quantum dots such as the frequency and illumination dependent peak height, a thermal shift of the tunneling resonances and the occurrence of the additional nonequilibrium peaks.
\end{abstract}

\maketitle

\section{Introduction}
On the road towards quantum information processing units, self-assembled InAs quantum dots are promising candidates. Their level structure resembles a paradigmatic quantum system - an artificial atom with discrete energy levels and good coherence properties~\cite{prechtel_decoupling_2016,drexler_spectroscopy_1994,michler_quantum_2000,shields_semiconductor_2007}. Recently, transform limited single photon emission from such a quantum dot was demonstrated~\cite{kuhlmann_transform-limited_2015}. Due to advances in the material quality~\cite{ludwig_ultra-low_2017} Stark shift control enables the stabilization of the emission line~\cite{prechtel_frequency-stabilized_2013} while insights in the level structure are gained by optical methods like photo-luminescence spectroscopy~\cite{ediger_peculiar_2007} and resonance fluorescence~\cite{kuhlmann_dark-field_2013}.

Due to the nearly ideal situation of an artificial atom in a solid state matrix, experiments on level degeneracy and carrier-carrier interaction are at hand. A new method of equilibrium states spectroscopy was established by applying transport spectroscopy of a quasi two-dimensional electron gas (2DEG) tunnel-coupled to quantum dots~\cite{marquardt_using_2009}. Utilizing the advantages of this method, even the degeneracy of the quantum states~\cite{beckel_asymmetry_2014} and existence of nonequilibrium states were successfully probed~\cite{marquardt_transport_2011}. However, magnetic dispersion measurements that allow one to further investigate the properties of quantum dot states were not yet feasible as the influence of the magnetic field on the 2DEG is too strong~\cite{zhou_tuning_2015}.

Capacitance-voltage spectroscopy is typically applied to (n-i-Schottky)-diode structures containing quantum dots which are tunnel coupled to a degenerately doped back contact to gain insight into the level structure and the Coulomb interaction of charge carriers in the quantum dots~\cite{drexler_spectroscopy_1994,warburton_coulomb_1998}. While some dynamics can be observed upon increasing the AC-probe frequency~\cite{luyken_dynamics_1999}, it is generally still a rather slow method where the Fermi level of the back contact is in equilibrium with the quantum dot ``charge'' levels and, thus, unable to observe any nonequilibrium processes. Labud et al.~\cite{labud_direct_2014} developed a sample design expanding the method to probe the Coulomb interaction of electrons and illumination induced metastable holes in the quantum dots. The holes in the quantum dots are leftovers from illumination induced electron hole pairs. Some of these excitons get dissociated in the electric field of the Schottky diode or they can tunnel to the back contact without prior recombination.
This way, up to 5-fold positively charged excitonic states could be observed by C(V) spectroscopy~\cite{labud_direct_2014}. Here, electrons tunnel from a three dimensional back contact into the quantum dots ground state while they are attracted due to Coulomb interaction with nonequilibrium holes stored in the quantum dots. Although the holes are certainly in metastable states, the spectroscopy is done in a quasi-equilibrium, as the electrons tunnel into the lowest possible state as soon as an oscillating gate voltage is applied, enabling resonant tunneling and further the elimination of stored holes by recombination.

Brinks et al.~\cite{brinks_thermal_2016} measured the electron tunneling process in this quasi-equilibrium situation and developed a corresponding model, but it has not yet been possible to achieve nonequilibrium tunneling in C(V) spectroscopy at low temperatures.
In principle, the illumination generated holes should be able to annihilate stored electrons and thus enable nonequilibrium state tunneling events. Until now a strong tunnel coupling of the former devices suppressed this mechanism.
One way to generate sufficient nonequilibrium states is to strongly increase the hole generation rate by extensive illumination.
A drawback of this method are light induced leakage currents through the device, which make C(V)-spectroscopic investigations impossible. As an alternative way, we increase the tunnel barrier length in order to reduce the electron tunneling rate to a level comparable to or even below the hole generation rate.
We will show that the generated holes have the ability to eliminate electrons in the quantum dots, leaving the dot in a nonequilibrium state. Subsequently, this state can be filled by an electron tunneling from the back contact. As the tunnel probability is highest for resonant tunneling (where the Fermi-level of the back contact matches a level of the quantum dot) additional peaks occur. We identify these peaks with electrons tunneling into nonequilibrium $s$-, $p$-, and $d$- states which are usually inaccessible via the slow DC-sweep in our C(V) spectroscopy. 
While the simple model of Brinks et al.~\cite{brinks_thermal_2016} describes the observed quasi-equilibrium C(V)-measurement features, we expand this to an extended theoretical model including electron elimination by holes, offering an excellent quantitative agreement with the observed features in the measurements.

The paper is organized as follows. In Sec.~\ref{sec:experiment} we present our experimental findings and provide an interpretation within the framework of the simple model of Warburton et al.~\cite{warburton_coulomb_1998}. The theoretical framework based on a master equation approach is discussed in Sec.~\ref{ssec:ME}. It is illustrated with a simple example in Sec.~\ref{ssec:example} and used to model the experiment in Sec.~\ref{ssec:modelling}. We provide conclusions in Sec.~\ref{sec:conclusion}.

\section{\label{sec:experiment}Experiment and Interpretation}
\begin{figure}
	\includegraphics[width=\columnwidth]{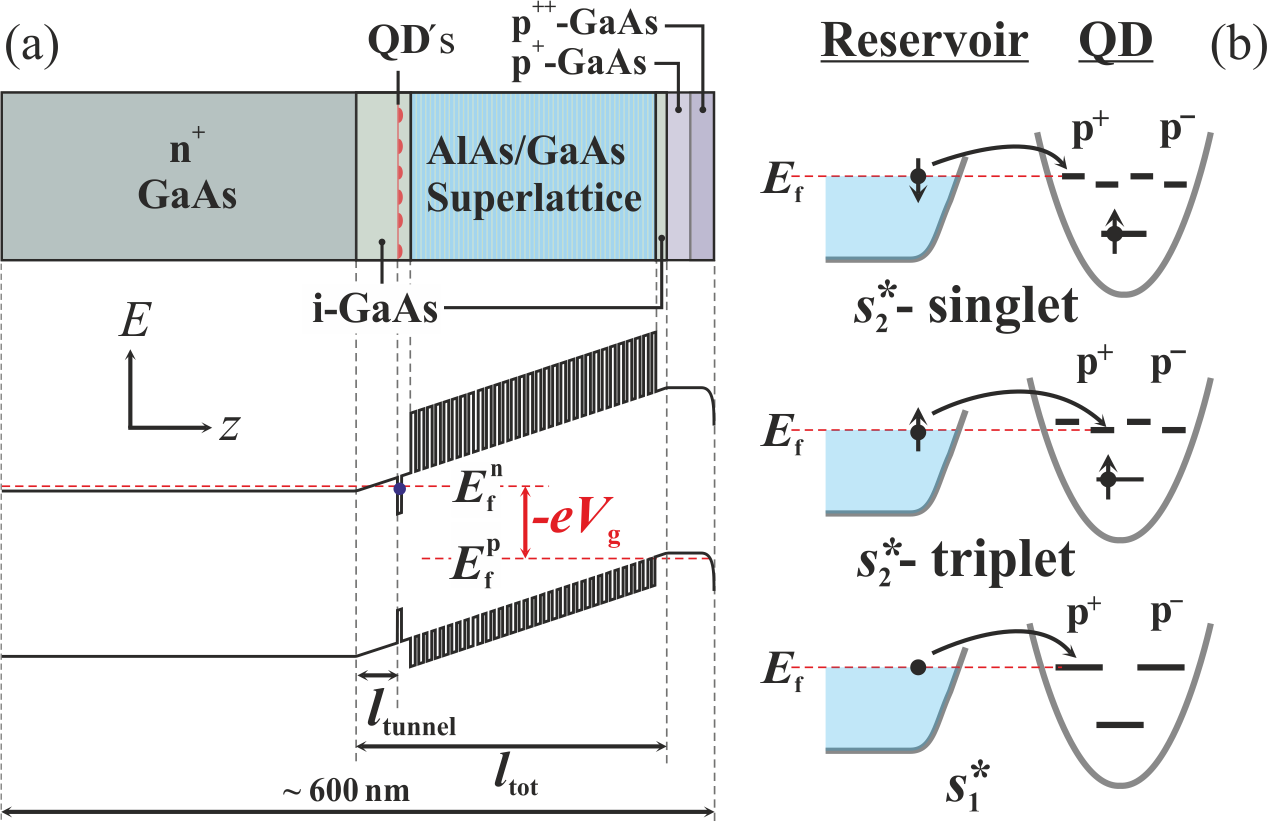}
	\caption{(a) Band structure sketch of the sample. (b) Electron tunneling into excited states. The quantum dot energy for resonant tunneling is ascending from top to bottom.}
	\label{fig:sample_structure}
\end{figure}

We study self-assembled InAs quantum dots in a GaAs based n-i-p-diode by capacitance-voltage spectroscopy under constant illumination. The quantum dots are tunnel coupled via a $\unit[35]{nm}$ GaAs tunnel barrier to a degenerately silicon doped ($N_\text{D}=\unit[2\times10^{18}]{cm^{-3}}$) GaAs back contact and capped by $\unit[11]{nm}$ GaAs followed by a blocking barrier, which consists of 50 periods of an AlAs/GaAs ($\unit[3]{nm}$/$\unit[1]{nm}$) short-period superlattice (SPS). An epitaxial, complementary-doped and semitransparent electrostatic gate~\cite{pal_ultrawide_2015,prechtel_decoupling_2016} is grown on top of the sample. It is composed of a $\unit[25]{nm}$ thick bulk carbon-doped GaAs layer ($N_\text{A}=\unit[3\times10^{18}]{cm^{-3}}$), followed by 40 periods of carbon-delta-doped and $\unit[0.5]{nm}$ carbon-doped GaAs layers ($N_\text{A}=\unit[1\times10^{19}]{cm^{-3}}$), see Fig.~\ref{fig:sample_structure}(a).
The wafer is processed by standard wet chemical etching to a mesa of $\unit[300\times300]{\mu m^2}$. Ohmic n-contacts are fabricated by indium solder to the corners of a $\unit[4\times5]{mm^2}$ sample, glued by silver paint to a chip carrier. Due to the high surface doping, we achieve ohmic p-contacs simply by wire bonding the semiconductor surface.
The measurements are performed in an insert of a liquid helium vessel at $\unit[4.2]{K}$.

The self-assembled quantum dots are "as grown" dots, i.e. their ground state emission wavelength in photoluminescence characterization at $T=\unit[77]{K}$ is centered around $\unit[1200]{nm}$. As we typically observe a shift of less than $\unit[5]{nm}$ reducing the temperature to the measurement temperature of $\unit[4.2]{K}$, we assume this $\unit[1.04]{eV}$ to be the ground state transition energy of the quantum dots in the experiment. 

Hole generation in the quantum dots is performed via IR-LED illumination inside the cryostat. The LED has an emission spectrally centered around $\unit[950]{nm}$ at $\unit[293]{K}$ and around $\unit[920]{nm}$ at $\unit[77]{K}$. We do not expect this to shift much for even lower temperatures. The photons generate electron-hole pairs inside the wetting layer or even in the quantum dots themselves. While a certain share of these pairs will directly recombine, some separate due to the internal electrical field or by electron tunneling into empty states in the back contact. Corresponding holes drop into or remain trapped inside the quantum dots. The according hole generation process in the quantum dots is expected to be proportional to the LED current~\cite{labud_direct_2014}.  Brinks et al. found a hole generation rate per quantum dot of $\unit[400]{Hz}$ for an LED current of $\unit[2]{mA}$ ~\cite{brinks_thermal_2016}.

Capacitance-voltage spectroscopy is measured by using an analogue adder, superimposing an AC voltage of $V_\text{AC, rms}=\unit[10]{mV}$ and a DC gate voltage $V_\text{g}$ on the epitaxial gate and measuring the AC current through the device at the n-contact at the same frequency $\omega/(2\pi)=\unit[92.3]{Hz}$ as the AC voltage with a Lock-in amplifier.
The $-90^{\circ}$ phase-shifted current is then converted to the differential capacitance $C(V_\text{g})$.

A common method to convert the gate voltage $V_\text{g}$ to an energy scale $E$ is to apply a simple geometric lever $\alpha=l_\text{tunnel}/l_\text{tot}$ relating the tunnel length $l_\text{tunnel}$ to the distance between the gate electrode and the back contact $l_\text{tot}$ ~\cite{lei_probing_2008},

\begin{equation}
	E=e\alpha(V_\text{built-in}- V_\text{g}),
\end{equation}
where $eV_\text{built-in}=\unit[1.52]{eV}$ is the band gap of our n-i-p-diode.

\begin{figure}
	\includegraphics[width=\columnwidth]{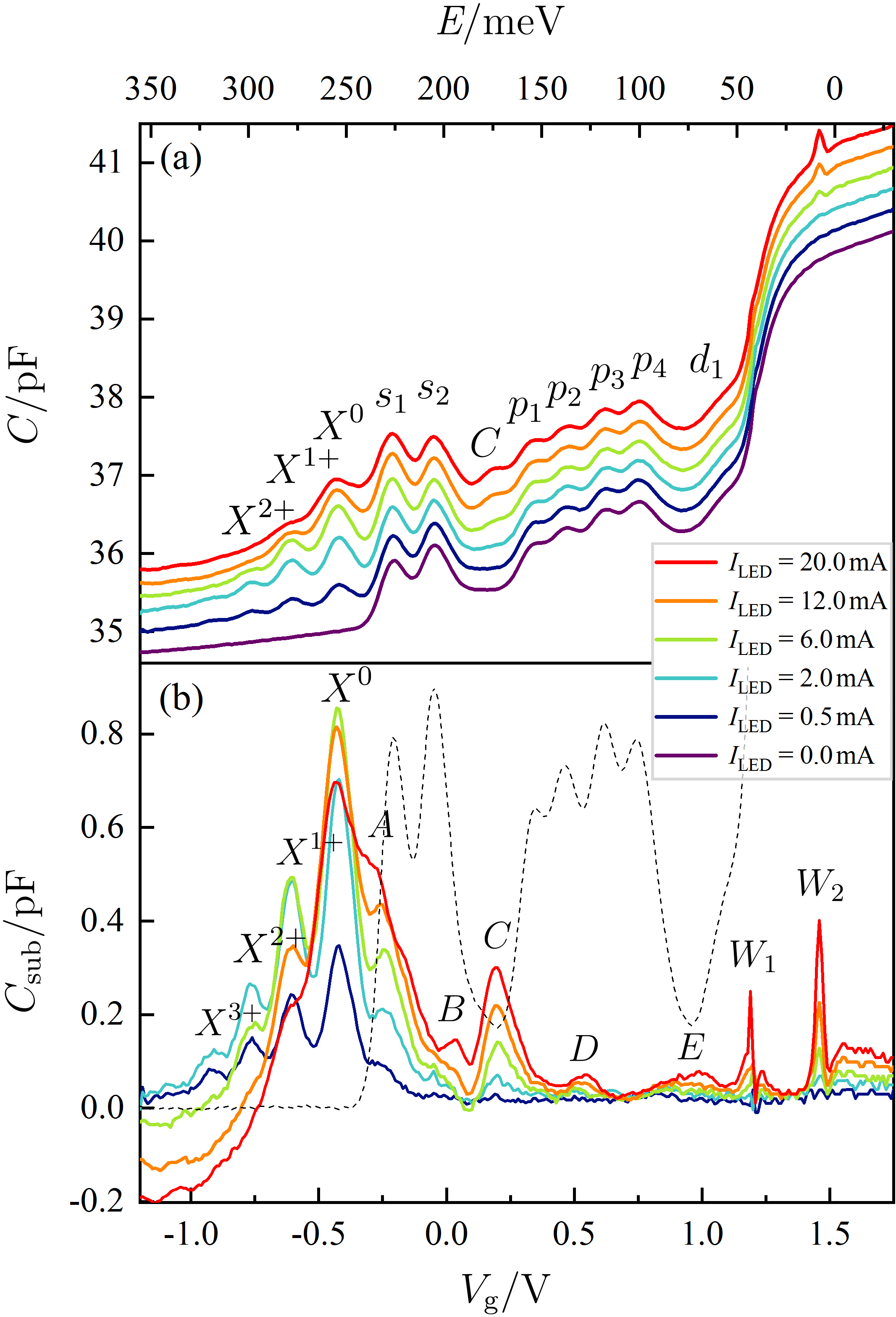}
	\caption{(a) Reference C(V) curve without illumination (violet) followed by C(V) curves under illumination at ca. $\unit[920]{nm}$ using different LED currents ranging from $\unit[0.5]{mA}$ up to $\unit[20]{mA}$. The C(V) spectra are offset by $\unit[0.25]{pF}$ for better visibility. (b) To reveal the illumination induced additional capacitances we subtracted the unilluminated reference curve from the illuminated ones. 
For a better allocation of the illumination induced peaks within the quantum dot spectrum, we inserted the unilluminated reference curve (dashed line) subtracted by an alleged background capacitance of the diode, using the almost linear part before the s-peaks for fitting.
The gate voltage is converted to energy by applying the simple lever-arm approach ~\cite{lei_probing_2008}.
}
	\label{fig:Peaks_LED_current_dependence}
\end{figure}

Apart from the standard peaks for tunneling into $s$- and $p$-states, under illumination exitonic features appear in the voltage range $V_\text{g}=\unit[-1]{V}$ to $V_\text{g}=\unit[-0.4]{V}$ which have already been reported by Labud et al.~\cite{labud_direct_2014}. Furthermore, a clear additional peak between the $s$- and the $p$-states appears as seen in Fig.~\ref{fig:Peaks_LED_current_dependence}(a) at $V_\text{g}=\unit[0.2]{V}$.
To reveal new peaks more clearly, we subtracted the nonilluminated reference spectrum from the illuminated ones in Fig.~\ref{fig:Peaks_LED_current_dependence}(b). A series of new peaks labeled A to E becomes visible. We assign these to nonequilibrium state electron tunneling into the quantum dots as discussed in detail below. We assume the sharp features $\text{W}_{1}$ and $\text{W}_{2}$ are related to wetting layer state tunneling.
Beside these new peaks, we also observe that the exitonic peaks reach a maximum at $\unit[6]{mA}$ for $X^{0}$ and $X^{1+}$ and $\unit[2]{mA}$ for $X^{2+}$ and $X^{3+}$ and decrease for higher LED currents.

\begin{figure}
	\centering
	\includegraphics[width=\columnwidth]{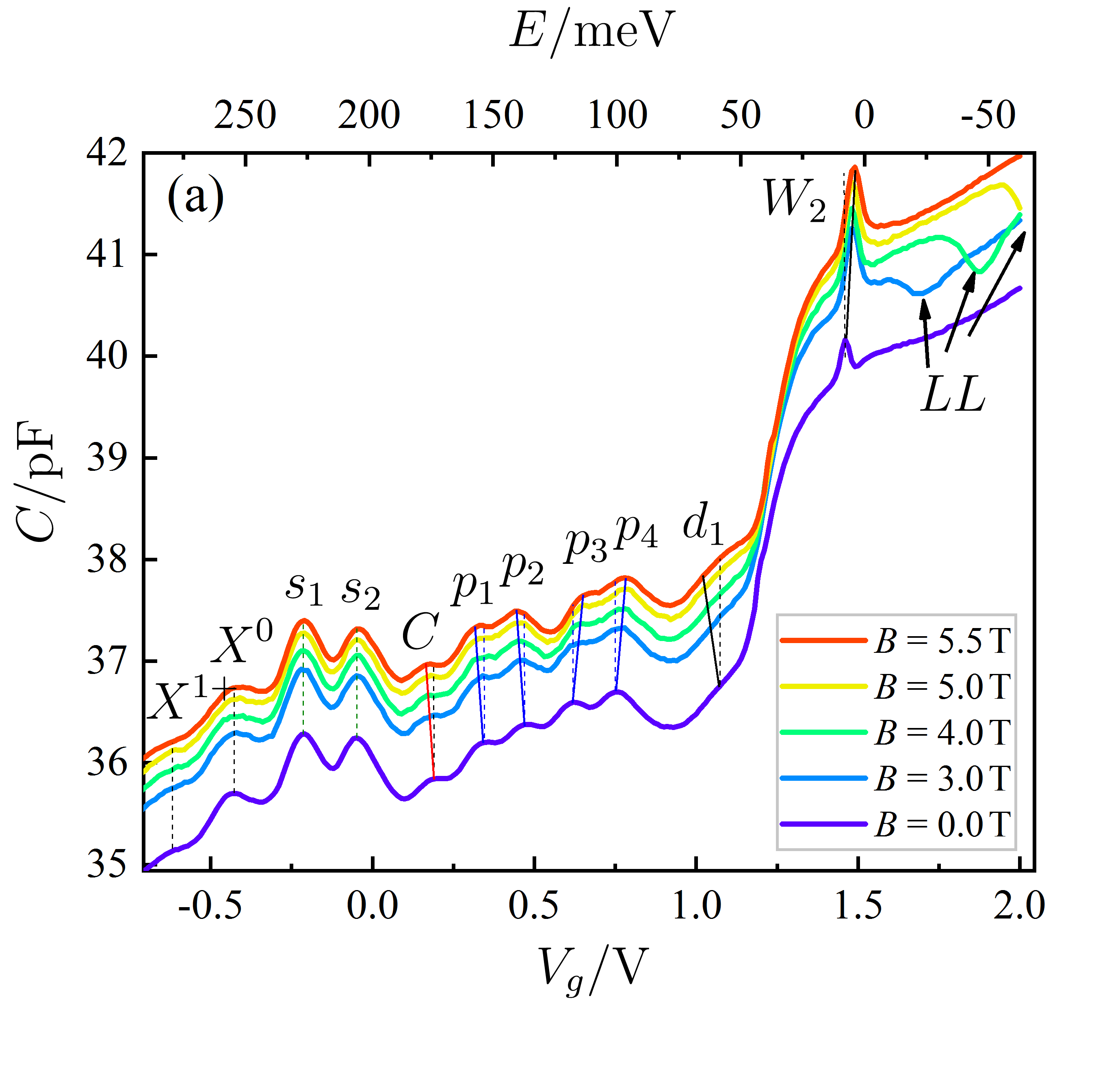}
	\hspace{0.05\textwidth}
	\includegraphics[width=\columnwidth]{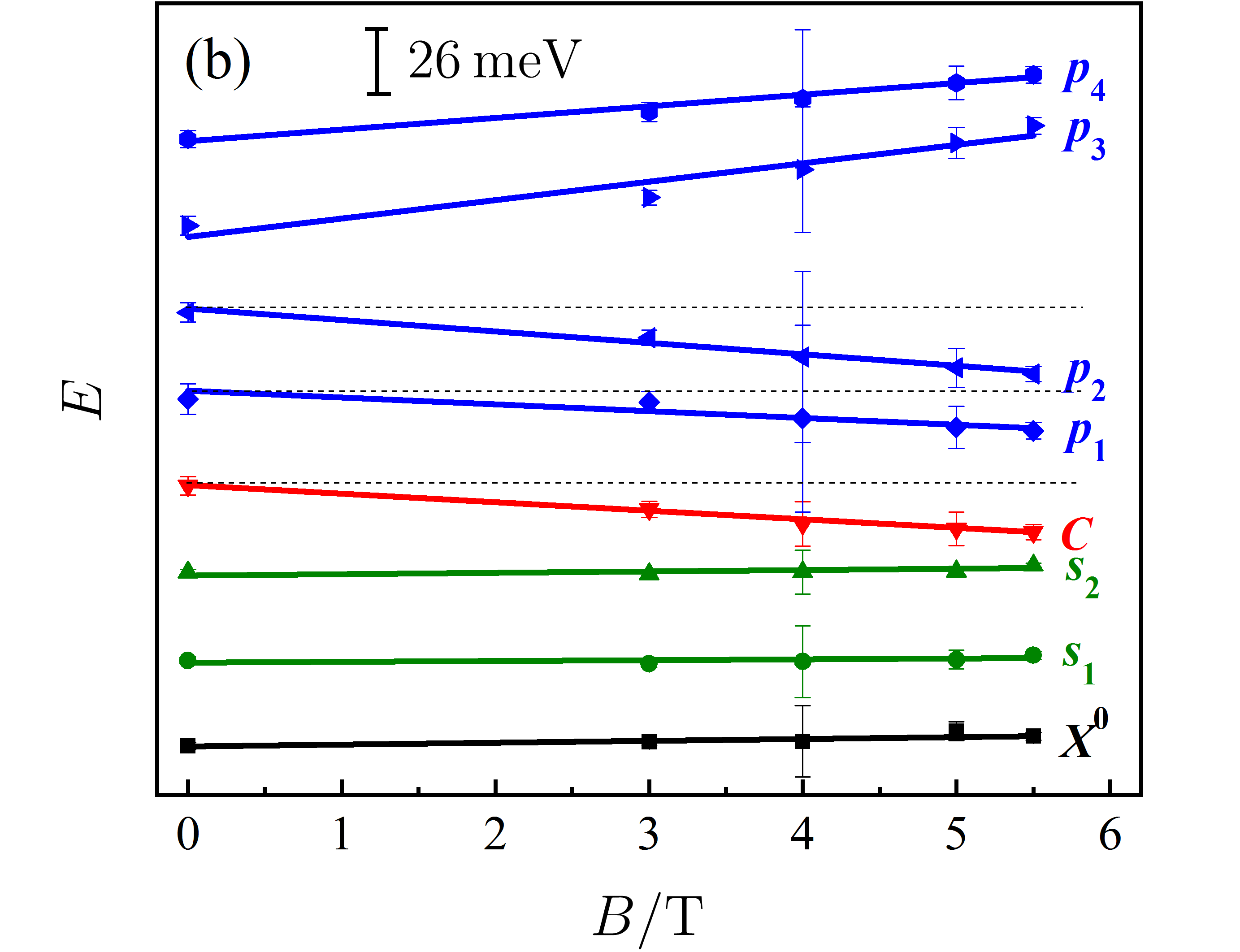}
	\caption{(a) C(V) spectrum with an illumination of $I_{LED}=\unit[20]{mA}$ and magnetic fields of up to $\unit[5.5]{T}$. The graphs are offset proportional to the magnetic field for clarity. Dashed lines are vertical; the solid lines are a guide to the eye for the shift due to the magnetic dispersion.
The dip labeled $LL$ corresponds to capacitive charging of Landau levels in the wetting layer~\cite{pal_probing_2016}.
 (b) Position of the different peaks and linear fits. The dashed lines correspond to dashed vertical lines in (a). For better visibility of the shifts, the peaks are offset. A clear shift of addition peak C is visible.}
	\label{fig:Magnetic_field_dispersion}
\end{figure}

In order to identify the observed peaks, we apply a magnetic field perpendicular to the sample plane.
The result plotted in Fig.~\ref{fig:Magnetic_field_dispersion} resembles the findings of  Labud et al. Apart from a diamagnetic shift, there is no significant shift of the $s$-states or the excitonic feature $X^{0}$ visible, but a clear linear shift occurs for the $p$-peaks~\cite{labud_direct_2014}.
The well resolved additional light-induced peak C between the $s$- and $p$-states shows a slope comparable to the $p$-state slopes. 
We thus assign this peak to a nonequilibrium $p$-state that we label $s_{2}^{*}$ in the following. 
The peak $W_{2}$ found during charging of the wetting layer shifts with a slope comparable to that of the $p$-peaks indicating that the effective masses of the involved charge carriers are alike. As $W_{1}$ and $W_{2}$ are much narrower than the inhomogeneously broadened quantum dot peaks, we consider that these peaks originate from tunneling into two-dimensional states, most probably exciton bound states in the wetting layer. The dip denoted $LL$ corresponds to the capacitive charging of Landau levels in the wetting layer, which is also visible without illumination~\cite{pal_probing_2016}.

\begin{table}
	\caption{\label{tab:peak_positions}Voltages and corresponding energies of the experimentally observed peaks as well as energies calculated after the simple model in~\cite{warburton_coulomb_1998} with $\hbar\omega_e=\unit[46]{meV}$, $\hbar\omega_h=\unit[25]{meV}$, and  $E^C_{ss}=\unit[21]{meV}$.}
	\begin{ruledtabular}
	\begin{tabular}{ldcc}
		Peak & V_{\textrm{g}}/\unit{V} & $E_\text{exp}/\unit{meV}$ & $E_\text{th}/\unit{meV}$ \\
		\hline
		$X^{1+}$ & -0.602 & 277 & 275\\
		$X^0$ & -0.425 & 254 & 251\\
		A ($X^{1+*}$) & -0.242 & 230 & 222\\
		$s_1$ & -0.209 & 226 & 226\\
		$s_2$ & -0.050 & 205 & 205\\
		B ($X^{0*}$) & 0.035 & 194 & 201\\
		C ($s_2^*$) & 0.194 & 173 & 170\\
		$p_1$ & 0.345 & 154 & 154\\
		$p_2$ & 0.466  & 138 & 143\\
		D ($s_1^{**}$) & 0.533 & 130 & 135\\
		$p_3$ & 0.616 & 118 & 125\\
		$p_4$ & 0.743 & 102 & 114\\
		E ($s_1^{***}$) & 0.940 & 75 & 89
	\end{tabular}
	\end{ruledtabular}
\end{table}

To assign the additionally found peaks, we use the simple model of Warburton et al.~\cite{warburton_coulomb_1998}, cf. Table~\ref{tab:peak_positions}. The model is based on the following key assumptions: (i) The vertical confinement in the dot is much stronger than the lateral confinement. Hence, the quantum dots are effectively two-dimensional. (ii) The two-dimensional confinement potential for electrons and holes is parabolic with confinement energies $\hbar\omega_e$ and $\hbar\omega_h$, respectively. The lowest single-particle states are labelled as $s$, $p$ and $d$, have degeneracy 2, 4 and 6, and energies $\hbar\omega_{e,h}$, $2\hbar\omega_{e,h}$ and $3\hbar\omega_{e,h}$, respectively. (iii) Quantization energies are larger than Coulomb interactions which allows for a perturbative treatment of the latter.

The Coulomb energy between two electrons in the $s$ state is given by 
\begin{equation}
	E^C_{ss}=\frac{e^2}{4\pi\varepsilon_0\varepsilon_r}\sqrt{\frac{\pi}{2}}\frac{1}{l_{e}},
\end{equation}
where
\begin{equation}
	l_e=\sqrt{\frac{\hbar}{m_e^*\omega_e}},
\end{equation}
with the effective mass $m_e^*$. A similar expression holds for the Coulomb repulsion between two holes in the $s$-state. The Coulomb energy between an electron in the $s$- and the $p$-state is given by $E^C_{sp}=3/4 E^C_{ss}$. The Coulomb attraction between an electron and a hole in the $s$-state is $E^{eh}_{ss}=\sqrt{2/[1+(l_h/l_e)^2]}E^C_{ss}$ while the Coulomb attraction between an electron in the $p$-state and a hole in the $s$-state reads $E^{eh}_{sp}=(2l_e^2+l_h^2)/(l_e^2+l_h^2)E^{eh}_{ss}/2$. Finally, the exchange interaction between an electron in the $s$- and $p$-states is given by $E^\text{ex}_{sp}=E^C_{ss}/4$.

The total energy of a given dot state with $N_e$ electrons and $N_h$ holes consists of the sum of all single-particle energies, the corresponding Coulomb terms, and the electrostatic energy arising from the coupling to the gate, $E^\text{tot}=E^\text{sp}+E^C-(N_e-N_h) e \alpha (V_\text{g}- V_\text{built-in})$. The excitation energies at which transitions between different dot states occur are given by the differences between the corresponding total energies.

As a concrete example, let us consider the energy of the $s_2^*$ state where one electron is in the $s$- state while a second electron occupies a $p$- state. The sum of the single particle energies is hence given by $E^\text{sp}=3\hbar\omega_e$. Apart from the Coulomb repulsion $E^C_{sp}=3/4 E^C_{ss}$ between the electrons, there is also the exchange energy $E^\text{ex}_{sp}=E^C_{ss}/4$ that the electrons gain if their spins are aligned parallel. This implies that the $s_2^*$ actually is split into two, a singlet state with total energy $E^\text{tot}=3\hbar\omega_e+3/4E^C_{ss}-3 e \alpha (V_\text{g}- V_\text{built-in})$ and a triplet state with total energy $E^\text{tot}=3\hbar\omega_e+1/2E^C_{ss}-3 e \alpha (V_\text{g}- V_\text{built-in})$. In the C(V) spectra the triplet state dominates over the singlet state because the charge on the dot is changed as soon as tunneling of a second electron onto the dot becomes energetically possible.

Using the simple model, we arrive at the following identification of the nonequilibrium peaks, cf. also Table~\ref{tab:peak_positions}. Peaks A ($X^{1+*}$) and B ($X^{0*}$) correspond to an electron tunneling into the $p$-state when the quantum dot is occupied with one electron and two or one holes, respectively, in the $s$-state.
Peak C ($s_2^*$) is due to the tunneling of an electron into the $p$-state when the dot is occupied with a single electron in the $s$-state.
Finally, peaks D ($s_1^{**}$) and E ($s_1^{***}$) arise from electrons tunneling into the $d$- and $f$-state, respectively, when the dot is empty.

\section{\label{sec:theory}Theory}
In the following, we present a theoretical model of C(V) spectroscopy based on a master equation approach~\cite{beenakker_theory_1991}. Our theory provides a significant extension of the simple model of Brinks et al.~\cite{brinks_thermal_2016}. In particular, it explains the occurrence of electronic and excitonic peaks in the C(V) spectrum as well as their respective height. Furthermore, it captures the thermal shift of peak positions connected to the degeneracy of states~\cite{brinks_thermal_2016}. In addition, our model also describes and refines the frequency-dependent suppression of peaks which has been reported, e.g., in the work of Luyken~\cite{luyken_dynamics_1999}. Finally, our theory also explains the occurrence of nonequilibrium peaks.
\subsection{\label{ssec:ME}Master equation}
In order to model the C(V) spectra theoretically, we assume that charging effects between different quantum dots can be neglected. This is a reasonable approximation for a sufficiently low quantum-dot density. In consequence, the contribution of different quantum dots to the C(V) spectrum is independent and, hence, it is sufficient to consider a single quantum dot tunnel coupled to an electronic reservoir. Furthermore, we assume that all dots have identical properties. Therefore, the peaks in the C(V) spectra are only thermally broadened. In the experiment, there is an additional broadening arising from the distribution of dot parameters which can be taken into account by convoluting the theoretical spectra with a Gaussian.
Within our theoretical framework, the state of the system is characterized by the time-dependent probabilities $P_{n_e,n_h}(t)$ to find the quantum dot occupied with $n_e$ electrons and $n_h$ holes. We collect these probabilities in a vector $\vec P(t)$. 
Transitions from one dot state to another are possible via four different processes:
(i) Tunneling of an electron into and out of the quantum dot with the rate $g_\pm \Gamma_e f^\pm(E_f-E_i)$ where $g_\pm$ denotes the degeneracy factor of the final state, $\Gamma_e$ is the tunnel coupling strength while $E_f$ and $E_i$ are the initial and final dot energy. The probability to find an electronic state occupied is given by the Fermi distribution $f^+(E)$ while the probability to find it empty is given by $f^-(E)=1-f^+(E)$.
(ii) creation of a hole on the quantum dot with rate $\Gamma_h$,
(iii) electron-hole recombination with rate $\Gamma_\text{rec}$
(iv) relaxation from an excited to the ground state with rate $\Gamma_\text{rel}$.
In the following, we assume that $\Gamma_\text{rec}\gg\Gamma_{e,h}$. This implies that we can neglect states with both, a finite number of electrons \emph{and} holes and take into account transitions involving such states with an effective transition rate $(\Gamma_\text{rec}^{-1}+\Gamma_{e,h}^{-1})^{-1}\approx \Gamma_{e,h}$. In addition, we also assume that $\Gamma_\text{rel}\gg\Gamma_{e,h}$. In consequence, any transition of the system will always start from the ground state (but may end up in an excited state of the system). Collecting the different transition rates in the matrix $\vec W(t)$, we arrive at the master equation
\begin{equation}
	\frac{d}{dt}\vec P(t)=\vec W(t)\vec P(t).
\end{equation}
We assume that the level positions inside the dots are varied slowly in time by an oscillating gate voltage $V_\text{rf}$, i.e., we have
\begin{equation}
	\label{eq:epstime}
	\varepsilon\to\varepsilon-e\alpha V_\text{rf}e^{i\omega t}.
\end{equation}
We now expand both the occupation probabilities as well as the transition rates in powers of $V_\text{rf}$ such that
\begin{align}
	\vec P&=\vec P^{(0)}+\vec P^{(1)} V_\text{rf}+\mathcal O(V_\text{rf}^2),\\
	\vec W&=\vec W^{(0)}+\vec W^{(1)} V_\text{rf}+\mathcal O(V_\text{rf}^2).
\end{align}
Collecting terms of equal powers in $V_\text{rf}$, we obtain from the master equation the set of equations
\begin{align}
	0&=\vec W^{(0)}\vec P^{(0)},\\
	i\omega \vec P^{(1)}&=\vec W^{(1)}\vec P^{(0)}+\vec W^{(0)}\vec P^{(1)}.
\end{align}
The vectors are normalized such that $\vec e^T\cdot\vec P^{(i)}=\delta_{i0}$ where $\vec e^T=(1,1,\cdots)$. 
The time-dependent charge flowing into the gate is given by
\begin{equation}
	Q(t)=e\sum_{n_e,n_h}(n_e-n_h) P^{(1)}_{n_e,n_h}\alpha V_\text{rf},
\end{equation}
and the associated current flow is $I(t)=d Q(t)/dt$. We then obtain for the effective impedance of the quantum dot 
\begin{equation}
	Z^{-1}_\text{eff}=\frac{I(t)}{V_\text{rf}e^{i\omega t}},
\end{equation}
which can be decomposed as $Z_\text{eff}^{-1}=R^{-1}_\text{eff}+i\omega C_\text{eff}$ to yield the effective resistance and capacitance.
\subsection{\label{ssec:example}Simple example}
\begin{figure*}
	\centering
	\includegraphics[width=\textwidth]{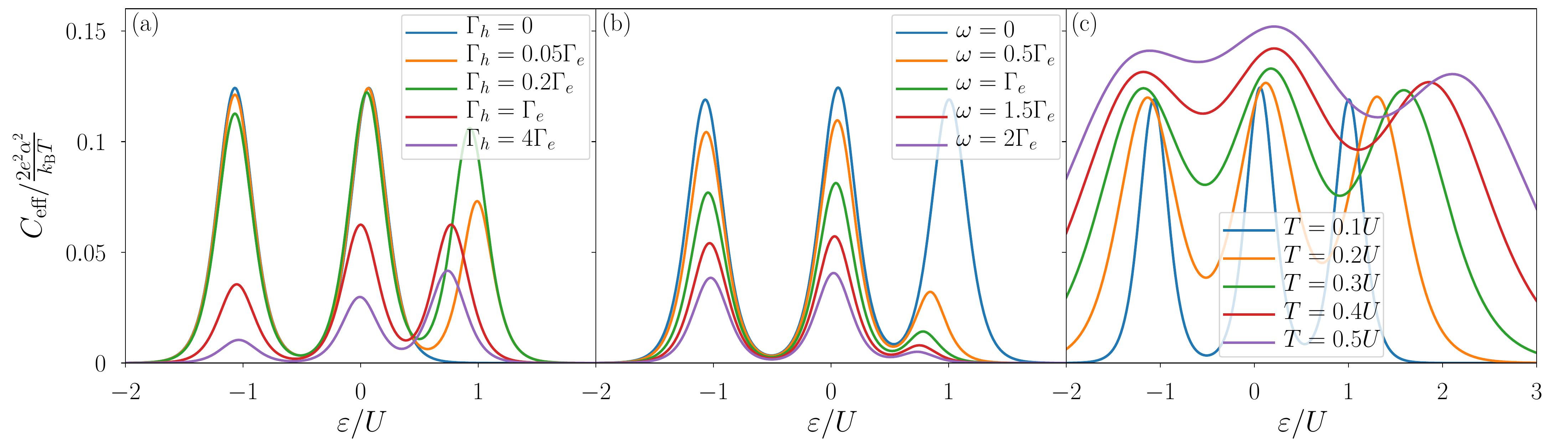}
	\caption{\label{fig:CVmodel} 
	Theoretical C(V) spectrum of a quantum dot with up to two electrons and one hole for (a) different values of the hole creation rate at $\omega=0.1\Gamma_e$, $T=0.1U$, (b) different driving frequencies at $\Gamma_h=0.1\Gamma_e$, $T=0.1U$, and (c) different temperatures at $\omega=0.001\Gamma_e$, $\Gamma_h=0.1\Gamma_e$. Other parameters are $U_{eh}=0.7U$.}
\end{figure*}
We now illustrate our general theoretical framework with the simple example of a quantum dot that can host at most two electrons and one hole in order to illustrate the key features of our theoretical model. The relevant dot states are the dot being empty, occupied with a single electron, doubly occupied and occupied with a single hole. In this basis, the matrix of transition rates reads 
\begin{widetext}
\begin{equation}
	\label{eq:W}
	\vec W=\left(\begin{array}{cccc} -2\Gamma_e f^+(\varepsilon)-\Gamma_h & \Gamma_e f^-(\varepsilon)+\Gamma_h & 0 & 2\Gamma_e f^+(\varepsilon-U_{eh}) \\ 2\Gamma_e f^+(\varepsilon) & -\Gamma_e f^-(\varepsilon)-\Gamma_e f^+(\varepsilon+U)-\Gamma_h & 2\Gamma_e f^-(\varepsilon+U)+\Gamma_h & 0 \\ 0 & \Gamma_e f^+(\varepsilon
	+U) & -2\Gamma_e f^-(\varepsilon+U)-\Gamma_h & 0 \\ \Gamma_h & 0 & 0 & -2\Gamma_e f^+(\varepsilon-U_{eh})\end{array}\right),
\end{equation}
\end{widetext}
where for simplicity we have suppressed the explicit time dependence of the level position $\varepsilon$, cf. Eq.~\eqref{eq:epstime}. In Eq.~\eqref{eq:W}, $U$ denotes the Coulomb energy that needs to be paid in order to occupy the dot with two electrons at the same time, and $U_{eh}$ is the attractive electron-hole interaction. While the effective capacitance of our simple example can be evaluated analytically, the resulting expressions are lengthy and therefore not given here.

In Fig.~\ref{fig:CVmodel} (a), we show the C(V) spectrum of the quantum dot for different values of the hole creation rate $\Gamma_h$. 
For vanishing $\Gamma_h$, there are two electronic peaks in the C(V) spectrum (blue line in Fig.~\ref{fig:CVmodel} (a)) which occur when the rate for tunneling in and tunneling out of electrons become equal to each other, i.e. at $\varepsilon=\kBT\log2$ for $s_1$ and $\varepsilon=-U-\kBT\log 2$ for $s_2$. At these positions, the charge on the dot changes gradually from empty to singly occupied and from singly occupied to doubly occupied.

At finite $\Gamma_h$, an additional peak develops when the hole-creation rate equals the tunneling-in rate of electrons, i.e., for $\Gamma_h=2\Gamma_e f^+(\varepsilon-U_{eh})$, indicating the formation of the neutral exciton $X^0$ on the dot. Interestingly, the excitonic peak height first increases with $\Gamma_h$, reaches a maximum at $\Gamma_h\approx 0.2\Gamma_e$ and then decreases again. This is because for very large $\Gamma_h$, hole creation dominates over electron tunneling, the dot spends most of the time in the state with $n_e=0$ and $n_h=1$ and, hence, the change of the dot charge across the peak becomes suppressed.
It is for the same reason that the height of the two electronic peaks also decreases when the hole creation rate is increased.

In Fig.~\ref{fig:CVmodel}(b), the suppression of peak height with frequency can be seen. We note that the height of the excitonic peak is more strongly suppressed than the height of the electronic peaks. At the same time, the position of the excitonic peak also shifts to smaller level positions upon increasing frequency while the position of the electronic peaks remains unaffected. So far, the frequency dependence of the excitonic peaks was investigated experimentally at low temperatures and small $\Gamma_h/\Gamma_e$ only where no such shift could be observed~\cite{labud_direct_2014}.
Quantitatively, we find that including the degeneracy of states modifies the frequency dependence of the peak height compared to the equivalent circuit model of Luyken et al.~\cite{luyken_dynamics_1999}. Considering tunneling between a single initial and final electronic state with degeneracies $g_+$ and $g_-$, respectively, we obtain a Lorentzian function
\begin{equation}
    \frac{C(\omega)}{C(\omega=0)}=\frac{4g_+^2g_-^2\Gamma_e^2}{4g_+^2g_-^2\Gamma_e^2+(g_++g_-)^2\omega^2}.
\end{equation}
We remark that our approach reproduces the results of the equivalent circuit model in the special case $g_+=g_-=1$. When considering tunneling into a twofold degenerate $s$-state, i.e. $g_+=2$ and $g_-=1$, we obtain a relative peak height of $16/25$ for a measurement frequency equal to the tunneling rate $\Gamma_e$ causing a correction to the value $1/2$ obtained when neglecting degeneracies~\cite{reuter_frequency-dependent_2004}.

Finally, in Fig.~\ref{fig:CVmodel} (c) we demonstrate the shift of peak positions with temperature. Apart from the linear shift of the electronic peaks mentioned above, there is a strong shift of the excitonic peak which is a direct consequence of the peak condition $\Gamma_h=2\Gamma_e f^+(\varepsilon-U_{eh})$. Both effects have been observed in experiment recently~\cite{brinks_thermal_2016}.

\subsection{\label{ssec:modelling}Modelling the experiment}
In order to reproduce the experimental findings within our master equation approach, we describe the quantum dot in the framework of the simple model of Warburton et al.~\cite{warburton_coulomb_1998}. We take into account states with up to six electrons on the dot as well as states with up to two holes. This allows us to capture peaks  in the C(V) spectra associated with tunneling into $s$- and $p$-states as well as with the formation of the neutral and positively charged exciton, $X^0$ and $X^{1+}$. As explained above, due to the fast electron-hole recombination rate, states with both, electrons and holes on the dot at the same time can be neglected when introducing associated effective transition rates. 
Due to the potential shape of the tunnel barrier in our experiments, tunnel couplings depend on energy. In order to model this effect, we assume different tunnel coupling strengths $\Gamma_{s,p}$ for tunneling into $s$- and $p$-states.
A more realistic treatment which models the energy dependency of tunneling rates via the WKB approximation~\cite{brinks_thermal_2016} would be required to obtain a quantitative agreement between the experimental and theoretical peak height which is beyond the scope of the present work.
Furthermore, degeneracy factors for tunneling in and out of different states are taken into account properly.

\begin{figure}
	\centering
	\includegraphics[width=\columnwidth]{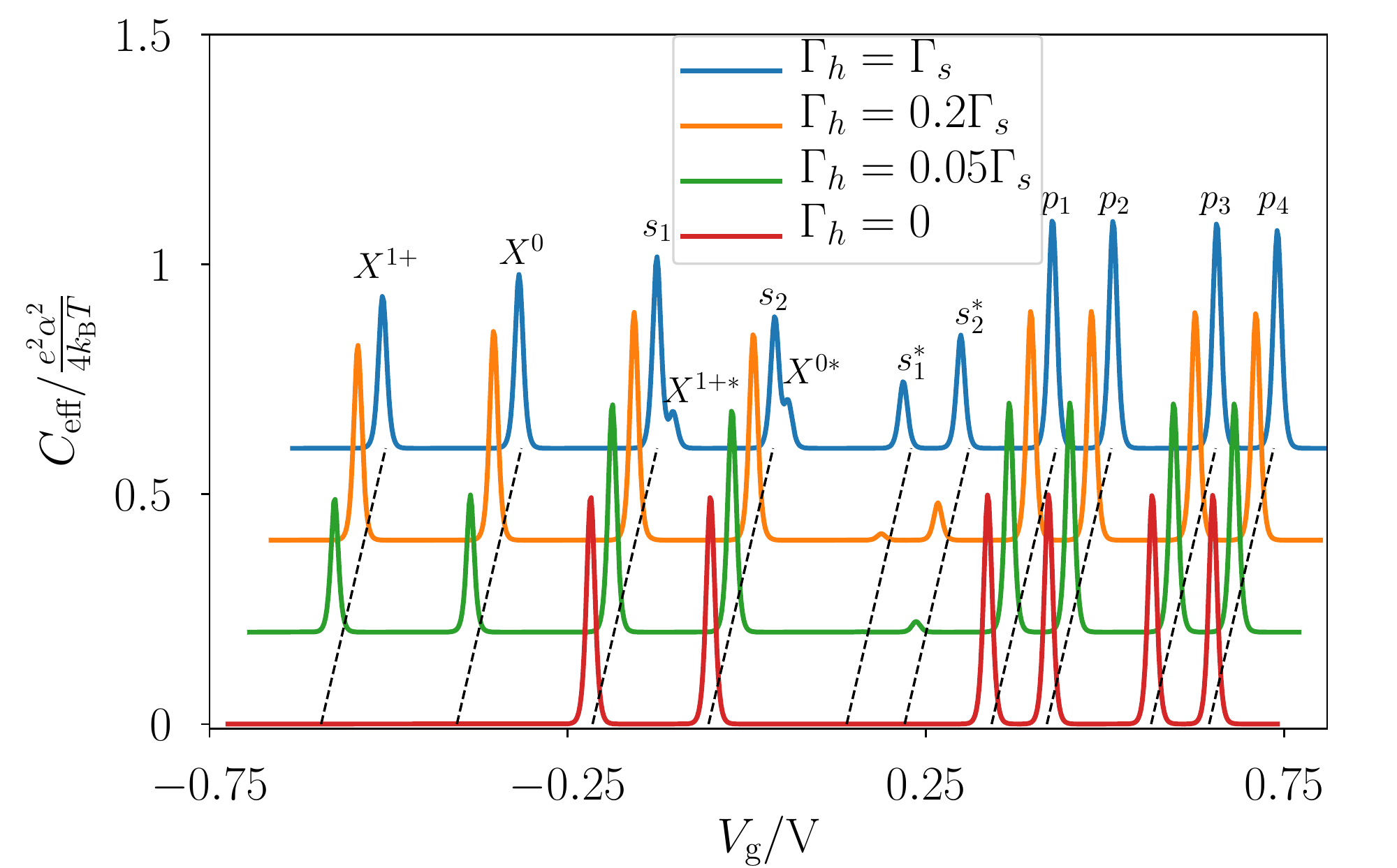}
	\includegraphics[width=\columnwidth]{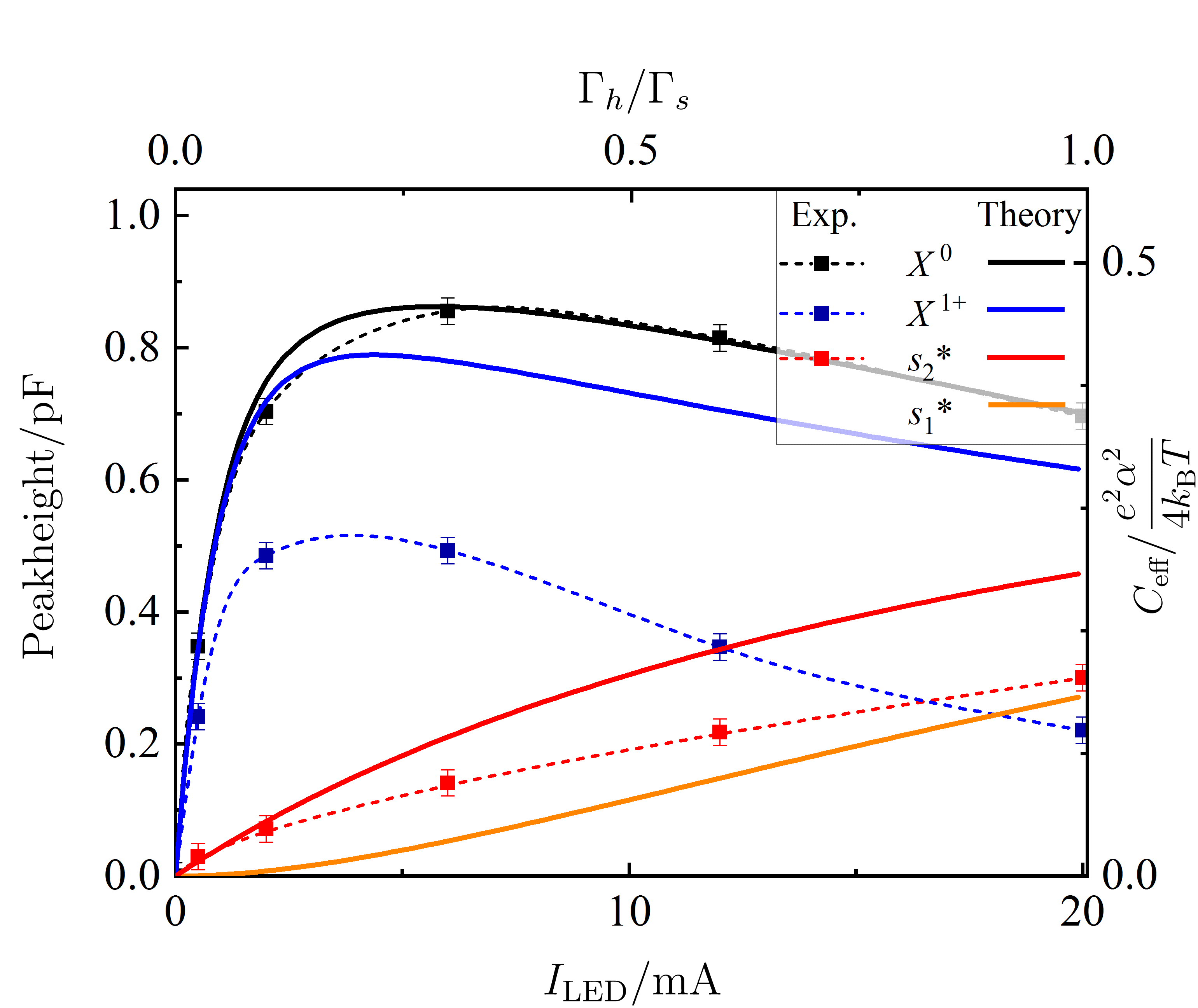}
	\caption{\label{fig:DensityPlots}
	(a) Modelled C(V) spectrum as function of the gate voltage for different values of $\Gamma_h$. Curves are offset and shifted for clarity. Dashed lines mark the peak positions. (b) Comparison of experimentally measured peak height as function of the LED current (left and lower axis, data points with dashed lines as guide to the eye) with the modelled peak height as function of the hole creation rate $\Gamma_h$ (right and upper axis, solid lines). Used model parameters: frequency $\omega=0.1\Gamma_s$, $\hbar\omega_e=\unit[46]{meV}$, $\hbar\omega_h=\unit[25]{meV}$,  $E^C_{ss}=\unit[21]{meV}$, $\kBT=\unit[0.5]{meV}$, $\Gamma_p=10\Gamma_s$.
	}
\end{figure}

The resulting C(V) spectrum is shown in Fig.~\ref{fig:DensityPlots} as a function of gate voltage and hole creation rate. At vanishing hole creation rate, we reproduce the usual peaks for tunneling into the $s$- and $p$-states of the quantum dot. Upon increasing the hole creation rate, two excitonic peaks, $X^0$ and $X^{1+}$ appear. 
When the hole creation rate becomes even larger $\Gamma_h\geq0.1\Gamma_e$, additional peaks due to the tunneling into excited states $s^*$ appear. They occur when a hole annihilates with an electron from the fully occupied $s$ shell and, subsequently, an electron tunnels into the $p$ shell. State $s_{1}^*$ corresponds to tunneling of an electron in the $p$-shell of an empty quantum dot, while $s_{2}^*$ corresponds to tunneling of an electron in the $p$-shell of a quantum dot already occupied by one electron in the $s$-shell. This can happen in an exited singlet or triplet configuration, cf. Fig.~\ref{fig:sample_structure}(b). The latter gains exchange energy from the parallel spin configuration and is thus energetically favoured. Due to the strong tunnel coupling of the $p$-shell, the energetically less favoured singlet configuration is not observed for the modelled $\Gamma_h$. As the hole creation rate is increased, these nonequilibriums become more and  more pronounced. Finally, at even larger hole creation rates $\Gamma_h>0.5\Gamma_e$, two additional peaks $X^{0*}$ and $X^{1+*}$ occur which are associated with tunneling  into excited excitonic states. Here, an electron tunnels into the $p$-states (instead of the $s$-state as compared to the excitonic ground state peaks) when the quantum dot is occupied with one or two holes. Once the hole creation rate becomes comparable to the rate of  tunneling into the $s$-states, the height of the excitonic peaks slowly decreases.

Comparing the results of our theoretical model with our experimental findings, we conclude that our model is able to reproduce the appearance of standard peaks in the C(V) spectra associated with tunneling into $s$- and $p$-states. Furthermore, our model captures the appearance of excitonic peaks under illumination as well as their slow decrease at large hole creation rates, cf. Fig.~\ref{fig:DensityPlots}(b). While this general trend is well reproduced, a quantitative agreement would require the inclusion of charging-related corrections to the tunnel rates in our theoretical modelling.
The model predicts a shift of the peaks at large hole creation rate and frequency at elevated temperatures.
In agreement with the experiment, this is not observed, if the temperature in the model is low.
Finally, our model also explains the appearance of nonequilibrium peaks associated with tunneling into excited electronic states if the theoretical peaks $s_1^*$ and $s_2^*$ are identified with the broad experimental peak $C$ as well as with tunneling into excited excitonic states if the theoretical peaks $X^{1+*}$ and $X^{0*}$ are identified with the experimental peaks $A$ and $B$, respectively.
\section{\label{sec:conclusion}Conclusion}
Nonequilibrium holes created by illumination enable electronic observation of excitonic states~\cite{labud_direct_2014}.
A quasi-equilibrium between hole generation and electron tunneling-in rate is created.
For weaker tunnel coupling (longer tunnel barriers), smaller electron tunneling-in rates enable the observation of resonant electron tunnel-in processes into excited states, experimentally confirmed by their magnetic dispersion. 
A quantitative model addresses the experimentally observed features.
Moreover predictions for C(V) measurements at elevated temperatures are made.
\acknowledgments
We acknowledge financial support from the Ministry of Innovation NRW. The work is financially supported by the BMBF Quantum communication program - Q.com-H 16KIS0109. The authors would also like to acknowledge the DFH/UFA CDFA-05-06 Nice-Bochum and RUB Research School. Additionally, A.D.W. and A.L. acknowledge support within the DFG SFB ICRC - TRR 160 Z1 project.


%

\end{document}